\documentclass[prl,preprint,letterpaper,showpacs,preprintnumbers,superscriptaddress,amsmath,amssymb]{revtex4}

\usepackage{graphicx}
\usepackage{dcolumn}
\usepackage{bm}

\begin{document}

\preprint{10/12/2003}

\title{Quantized charge pumping through a quantum dot\\
by surface acoustic waves}

\author{J. Ebbecke}
\affiliation{National Physical Laboratory, Queens Road, Teddington TW11 0LW, United Kingdom}
\affiliation{Cavendish Laboratory, University of Cambridge, Madingley Road, \\Cambridge CB3 0HE, United Kingdom}
 \author{N. E. Fletcher}
\author{T. J. B. M. Janssen}
\affiliation{National Physical Laboratory, Queens Road, Teddington TW11 0LW, United Kingdom}
\author{F. J. Ahlers}
\affiliation{Physikalisch-Technische Bundesanstalt, Bundesallee 100, D-38116 Braunschweig, Germany}
\author{M. Pepper}
\author{H. E. Beere}
\author{D. A. Ritchie}
\affiliation{Cavendish Laboratory, University of Cambridge, Madingley Road, \\Cambridge CB3 0HE, United Kingdom}

\date{\today}

\begin{abstract}
We present a realization of quantized charge pumping. A lateral quantum dot is defined by metallic split gates in a GaAs/AlGaAs heterostructure. A surface acoustic wave whose wavelength is twice the dot length is used to pump single electrons through the dot at a frequency $f=3~$GHz. The pumped current shows a regular pattern of quantization at values $I=nef$ over a range of gate voltage and wave amplitude settings. The observed values of $n$, the number of electrons transported per wave cycle, are determined by the number of electronic states in the quantum dot brought into resonance with the fermi level of the electron reservoirs during the pumping cycle.

\end{abstract}


\maketitle

The concept of quantized charge pumping was first suggested by Thouless \cite{Thouless83}. In an infinite one-dimensional channel with a gap in the excitation spectrum, non-interacting electrons are driven by a slowly moving periodic potential. The resulting dc current $I$ is quantized as $I=ef$, where $e$ is the electronic charge and $f$ the frequency of the moving potential. The electrons always remain in the ground-state and the mechanism is reminiscent of the Archimedean screw. In recent years, this idea has been extended to various finite systems in semiconductor nanostructures \cite{Hekking91,Altshuler99} and in particular charge pumps based on quantum dot (QD) devices have received considerable theoretical and experimental attention \cite{Brouwer98,Aleiner98,Shutenko00,Levinson01,Kouwenhoven91,Switkes99}. Accurate quantized charge pumping is of practical importance as a quantum standard for electric current \cite{Niu90}.

Quantized charge transport has been presented in metallic single electron tunneling (SET) \cite{Pothier92} and semiconductor QD \cite{Kouwenhoven91} devices where the system is in the Coulomb blockade regime. Phase-shifted AC-voltages are applied to surface gates to control the electron tunneling probability. Single electrons are transported through these devices at frequencies of roughly 10~MHz.

In this letter we report quantized charge pumping at a frequency of 3~GHz. A lateral QD is induced by metallic split gates in a GaAs/AlGaAs heterostructure and single electrons are pumped through the resonant states of this QD by the use of a surface acoustic wave (SAW). 
A key difference from the mechanism of the turnstile devices of Ref. \cite{Kouwenhoven91} and Ref. \cite{Pothier92} is that in our case as well as modulating the potential barriers of the QD, the traveling SAW also oscillates the electronic states within the QD relative to the Fermi level of the surrounding two-dimensional electron gas (2DEG).

This work is closely related to the single electron transport through one-dimensional channels by SAWs \cite{Shilton96, Talyanskii97, Janssen00, Ebbecke00}. Here, the combination of the dynamic potential due to the SAW and the static potential of the one-dimensional channel is thought to define traveling QDs, and the quantization of the current is a result of the Coulomb repulsion between electrons transported in these QDs. However, recent results have indicated that in these systems unintentional static QDs formed by impurity potentials could be involved in the transport mechanism \cite{Fletcher03}. 

\begin{figure}
  \begin{center}
	 \includegraphics[width=2.5in]{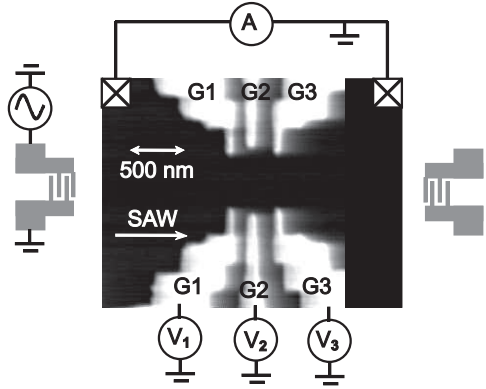}
	\end{center} 
	\caption{Schematic sketch of the measurement set-up with an electron microscope picture of the central region of the split-gates.}
	\label{fig:fig1}
\end{figure}
 
In Fig.~\ref{fig:fig1} a scanning electron micrograph of the central part of the sample is shown together with a schematic diagram of the measurement set-up. 
Three independent split gates are fabricated to induce a QD of 500nm length in the 2DEG located 90 nm below the sample surface. The 2DEG is a GaAs/AlGaAs heterojunction with electron density of $1.66~\times~10^{15}$~m$^{-2}$  and mobility of 128~m$^2$/Vs (both measured in the dark at T = 1.5~K). The two outer split gates define the entrance and the exit of the QD and are both 100~nm long with gaps of 600~nm. The middle gate is 100~nm long and has a gap of 800~nm. Two samples with identical split gate dimensions have been investigated. On sample $A$ there are two SAW transducers, on opposite sides of the split gates, with an interdigital period of 1~$\mu$m. This was chosen to give a SAW of wavelength approximately twice the QD length. The SAW velocity of 2800~m/s on GaAs gives an operating frequency of 2800~MHz. Sample $B$ has been processed with interdigital periods of 1~$\mu$m (2800~MHz) and 3~$\mu$m (980~MHz). 

When the two outer split gates were set to a conductance $G_{1,3} < 2e^2/h$ a series of equally spaced Coulomb blockade oscillations (CBO) was detected as the voltage on the middle gate, $V_2$, was swept close to conduction pinch-off. By varying the applied source-drain bias voltage, a charging energy of the induced QD of $E_C~\approx~$0.25~meV was deduced. At the temperature of 0.3~K of our measurements no distinct structure was observed in the CBO peaks and therefore we conclude that the difference between the quantum mechanical eigenstates of the QD is $\Delta\epsilon~\alt$~25~$\mu$eV~$\approx~kT$. 

\begin{figure}
	\includegraphics[width=3in]{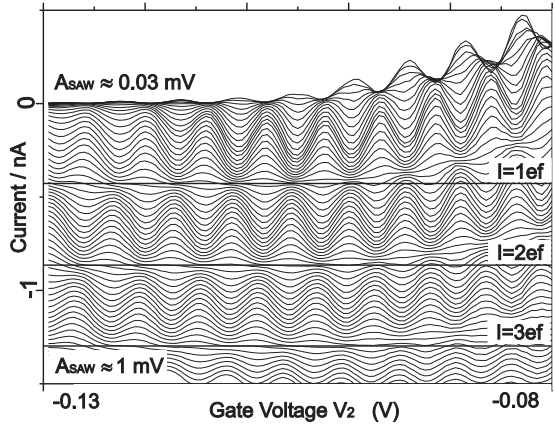}
	\caption{Acousto-electric current as a function of middle gate voltage, $V_2$, for fixed voltage $V_1$~=~-0.765~V and  $V_3$~=~-0.78~V. The rf power has been increased gradually from $P$ = -30~dBm (top) to $P$ = 0~dBm (bottom) at the resonance frequency of $f$~=~2697.05~MHz and source-drain voltage of $V_{SD}$ = 80~$\mu$V.}
	\label{fig:bottles}
\end{figure}

Fig.~\ref{fig:bottles} presents measurements of the acousto-electric current as a function of the voltage on the middle gate for sample $A$. The two outer split gates were held at a fixed voltage to define the entrance and exit tunnel barriers of the QD. The rf power applied to the SAW transducer was varied from $P_{SAW}$~=~-30~dBm (top curve) to $P_{SAW}$~=~0~dBm (bottom curve). The potential amplitude of the SAW resulting from these drive levels has been deduced by a method described in Ref.~\cite{Fletcher03}; 0~dBm measured at the generator corresponds to an amplitude $A_{SAW}~\approx$~1~mV at the QD. For the smallest SAW level ($P_{SAW}$~=~-30~dBm, $A_{SAW}~\approx$~0.03~mV), the current shows several CBO peaks before conductance-pinchoff, exactly as for no applied SAW. For larger amplitudes, the positive current due to the small applied source-drain bias changes to a negative current driven against this bias by the SAW. This acousto-electric current shows a very regular pattern which is aligned with both the CBO positions and the values of $I=nef$. For middle gate settings which give a peak in the CBO, the curves are most strongly grouped around odd values of $n$; for gate settings between peaks (maximum blockade), the curves tend to even values of $n$.


The piezoelectric potential of the traveling surface acoustic wave $A(x-{\omega}t)$ can be regarded as a superposition of two standing waves: $A(x-{\omega}t)=A_0~cos(2{\pi}t/T)sin(2{\pi}x/\lambda)+A_0~sin(2{\pi}t/T+{\pi})cos(2{\pi}x/\lambda)$ with $T$ the period of the pumping cycle and $\lambda$ the wavelength of the SAW. The dimensions of the device have been designed such that the wavelength $\lambda$ is double the length of the induced QD. The resulting phase difference $\phi=\pi$ in $sin(2{\pi}x/\lambda)$ and $cos(2{\pi}x/\lambda)$ at the entrance and exit potential barriers leads to a turnstile like modulation. As a result of the Coulomb blockade, a fixed number of electrons are transported during each SAW cycle. In contrast to the original turnstile mechanism presented in Ref. \cite{Kouwenhoven91} and Ref. \cite{Pothier92} our device does not require a source-drain bias to pump electrons through the QD. The SAW changes the defining potential of the static QD and therefore moves the single electron QD states energetically during the SAW cycle. This has the effect of producing a `local bias' across the tunnel barriers. In particular this SAW transport mechanism allows electrons to be pumped against an applied source-drain bias and can be regarded as a nanometer scale realization of an Archimedean screw.

The periodic pattern observed in Fig.~\ref{fig:bottles} can be explained by using the turnstile model and the assumption that the SAW amplitude provides a local bias. Changing the middle gate voltage brings different discrete electronic states of the QD into resonance with the Fermi levels of the source and drain 2DEG. At a CBO peak, one state is exactly aligned with the 2DEG, and the SAW transports one electron per cycle. At the gate voltages exactly in between the CBO peaks the Fermi levels of the source and drain 2DEG are in between two QD states. In this case the SAW amplitude needs to be increased to at least the level separation of the QD states. The SAW amplitude then encloses two QD states and therefore two electrons per SAW cycle are transported \cite{Fletcher03}.

\begin{figure}
	\includegraphics[width=3in]{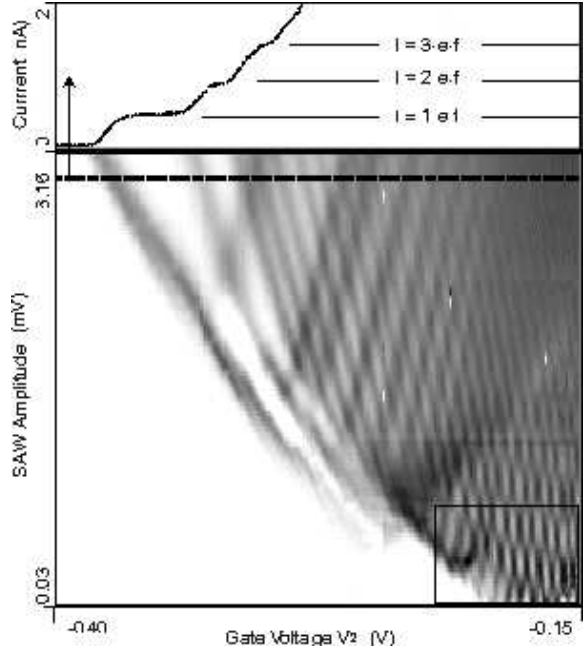}
	\caption{a) Numerical derivative of the acousto-electric current shown as a two-dimensional grey scale plot for a wide range of middle gate voltage, $V_2$, and SAW amplitude for sample $A$. White relates to a small value of the derivative and black to a rapid change of the acousto-electric current. The outer gates have been set at $V_{G1}$ = -0.74~V and $V_{G3}$ = -0.76~V and the SAW frequency is $f$ = 2697.05~MHz at $T$ = 500~mK. b) Current versus gate voltage along the line indicated in panel a).}
	\label{fig:fan}
\end{figure}

In Fig.~\ref{fig:fan}a) we show the acousto-electric current for a wide range of middle gate voltage, $V_{2}$, and SAW amplitude measured in sample $A$. Here the numerical derivative of the current with respect to applied SAW amplitude is plotted as a grey-scale map; the parameter space of Fig.~\ref{fig:bottles} is outlined in the lower right corner.
It can be seen that at least 18 different single-electron states of the QD have been brought into resonance by the SAW for different settings of $V_{2}$. After conduction pinch-off (the white lower left area of Fig.~\ref{fig:fan}a) the pattern shifts to progressively higher SAW amplitudes. This can be explained by the fact that the conductance of the two outer point contacts decreases for more negative $V_{2}$ and hence the SAW amplitude has to be increased in order to pump electrons through the QD. The precision of the current quantization is enhanced beyond conduction pinch-off because co-tunneling and leakage as a result of the applied source-drain bias are progressively suppressed. The line graph in Fig.~\ref{fig:fan}b shows a single acousto-electric current scan at a relatively high SAW amplitude to demonstrate the $I=nef$ sequence. 

Results very similar to those shown in Fig.~\ref{fig:fan}a, were obtained using the transducer on the opposite side of the QD; this transducer launches a SAW in the reverse direction, so the polarity of the pumped current is also reversed. 
A similar result was also obtained with sample $B$ using the 1~$\mu$m transducer. When using the second transducer on this sample to launch a SAW of 3~$\mu$m wavelength, the pattern of transitions remained largely unchanged; in this case, however, the acousto-electric current was less well quantized and found to be more sensitive to the applied source-drain bias.
For a SAW of 3~$\mu$m wavelength, the phase difference between the entrance and exit barriers in a 500~nm QD is $\pi/3$. Therefore, during part of the pumping cycle both barriers are relatively low and the current is sensitive to co-tunneling and leakage driven by the applied source-drain bias. The periodic pattern of current quantization is similar in all cases, as this is determined by the properties of the QD, which was induced by identical split-gates on both samples.


In summary we have realized quantized charge pumping mediated by surface acoustic waves at 3~GHz which is an increase in frequency of more than 2 orders of magnitude over the previously reported turnstile devices. The use of a SAW, rather than direct gate modulation, provides a simple means to interact with a QD at such high frequencies. Although the accuracy of the current quantization observed in our first devices is only of the order $10^{-3}$, this new system opens a promising route for the development of a quantum standard of current. Further experiments with smaller quantum dots should improve the accuracy. An interesting system to investigate in this context would be carbon nanotubes where a charging energy of more than 6~meV has been demonstrated \cite{Nygard01}.

\begin{acknowledgments}
We thank Miriam Blaauboer, Peter W{\"o}lfle and Piet Brouwer for useful discussions. F.J.A wants to thank the Glazebrook Foundation and the Deutsche Forschungsgemeinschaft for their support. This work was supported by the National Measurement System Policy Unit of the Department of Trade and Industry, UK
\end{acknowledgments}

\end{document}